\documentstyle[aps,eqsecnum,preprint,tighten]{revtex}

\begin{document}

\draft


\preprint{\vbox{Submitted to Phys.\ Rev.\ D\hfill DOE/ER/40762--027\\
                \null\hfill UMPP \#94--082}}

\title{Calculation of $\langle p|\overline{u}u-\overline{d}d|p\rangle$
from QCD sum rule and \linebreak the neutron-proton mass difference}
\author{Xuemin Jin and Marina Nielsen%
\thanks{Permanent address: Instituto de F\'{\i}sica,
Universidade de S\~ao Paulo, 01498 - SP- Brazil.}}
\address{Department of Physics and Center for Theoretical Physics\\
University of Maryland, College Park, Maryland 20742}
\author{J. Pasupathy}
\address{Center for Theoretical Studies, Institute of Science\\
Bangalore, India 560 012}
\date{\today}

\maketitle
\begin{abstract}
The proton matrix element of the isovector-scalar density, $\langle
p|\overline{u}u-\overline{d}d|p\rangle/2M_p$, is calculated by
evaluating the nucleon current correlation function in an external
isovector-scalar field using the QCD sum-rule method. In addition to
the usual chiral and gluon  condensates of the QCD vacuum, the response
of the chiral condensates to the external isovector field enters the
calculation.  The latter is determined by two independent methods.  One
relates it to the difference between the up and down quark chiral
condensates and the other uses the chiral perturbation theory.  To
first order in the quark mass difference $\delta m=m_d-m_u$, the
non-electromagnetic part of the neutron-proton mass difference is given
by the product $\delta m \langle
p|\overline{u}u-\overline{d}d|p\rangle/2M_p$; the resulting value is in
reasonable agreement with the experimental value.
\end{abstract}


\section{INTRODUCTION}
\label{intro}

Understanding the properties of the nucleon is of obvious importance in
hadron physics. A variety of nucleon matrix elements of bilinear quark
operators have been evaluated in the past using various approaches.
The QCD sum-rule method originally proposed by Shifman, Vainstein, and
Zakharov \cite{svz}  and its extension by Ioffe and Smilga
\cite{ioffe2} for  external field problems, provide a tractable
framework for the study of these nucleon matrix elements.  These
include the matrix element of electromagnetic current to determine the
magnetic moments \cite{ioffe2,balitsky1,chiu1}, the matrix element of
the axial vector current to find the renormalization of nucleon axial
coupling constant  \cite{belyaev2,chiu2}, the matrix element of the
quark part of the energy momentum tensor, which gives the momentum
fraction carried by the up and down quarks in deep inelastic scattering
\cite{kolesinchenko1,belyaev3}, and the matrix element of
isoscalar-scalar current for evaluating the nucleon Sigma term
\cite{jin1}.  In the present work, we calculate the matrix element of
the isovector-scalar density, $\langle p|\overline{u}u-
\overline{d}d|p\rangle/2M_p$, following the same external field
approach.

The appearance of the external field leads to specific new features in
QCD sum rules  which distinguish them from those in the absence of the
external field. Thus the phenomenological representation of the
correlation function written in terms of physical intermediate states
contains a double pole at the nucleon mass  whose residue contains the
matrix element of interest. In addition there are single pole terms
which arise from the transition matrix element between the ground state
nucleon and excited states. These later contributions are not
exponentially damped after Borel transformation relative to the double
pole term and should be retained in a consistent analysis of the sum
rules. In the theoretical side of the sum rules expressed in terms of
an operator product expansion (OPE) the external field contributes in
two different ways--by directly coupling to the quark fields in the
nucleon current and by polarizing the QCD vacuum.

Since for our problem the external field is a Lorentz scalar,
non-scalar correlators cannot be induced in the QCD vacuum. However the
external field does modify the quark and gluon condensates already
present in the QCD vacuum. It turns out that for the problem under
study the most important one is the response of the up and down quark
chiral condensates to the external field which can be described by a
susceptibility $\chi$. This can be determined by writing a spectral
representation, which can be evaluated for example by using chiral
perturbation theory. Alternatively, since in the real world up and down
quark masses are different, $\delta m = m_d - m_u $  itself can be
regarded as an external isovector scalar field and $\chi$ can be
related to the isospin breaking in quark condensates $\gamma\equiv
(\langle 0|\overline{d}d-\overline{u}u|0\rangle)/\langle
0|\overline{u}u|0\rangle$ and is given to first order in $\delta m $ by
$\chi = -\gamma / \delta m$.  At the current level of accuracy where
these two determinations can be done, they are mutually consistent.

In Sec.~\ref{sumrule} we derive the sum rules for the nucleon current
correlation function in an external isovector-scalar field and describe
the analysis of these sum rules, which leads us to determine $\langle
p|\overline{u}u-\overline{d}d|p\rangle/2M_p$.  We also study its
dependence on the $\chi$ value.

In Sec.~\ref{npd} we show that an evaluation of the matrix element
 $\langle p|\overline{u}u - \overline{d}d|p\rangle/2M_p$ enables us to
 compute the non-electromagnetic part of the neutron-proton mass
difference.  Since the 1970 's it has been recognized that the
empirical mass difference arises from two sources. One is purely
electromagnetic and yields a contribution of $-0.76\pm
0.30\,\text{MeV}$\cite{gasser1} to the neutron-proton mass difference.
The second source is the difference between up and down quark masses,
which is of the order of the quark masses themselves. This difference
leads to larger contribution which overrides the electromagnetic
contribution making neutron heavier than proton.  To first order in
$\delta m $ this later contribution is given by the product of
 $\delta m$ and $\langle p|\overline{u}u-\overline{d}d|p\rangle
/2M_p$.

Recently several authors \cite{hatsuda1,yang1,adami1,eletsky1} have
applied the QCD sum rules for the nucleon mass to extract the
neutron-proton mass difference by including the up and down quark
masses in the mass sum rules for proton and neutron and the isospin
breaking in the condensates. The relation of these works to ours is
described in Sec.~\ref{discussion}, where we also briefly comment on
the Nolen-Schiffer anomaly.


\section{Sum-rule calculation}
\label{sumrule}

The procedure that we use for calculating the matrix element $\langle
p|\overline{u}u-\overline{d}d|p\rangle/2M_p$ follows the same pattern
as used in Refs.~\cite
{ioffe2,balitsky1,chiu1,belyaev2,chiu2,kolesinchenko1,belyaev3,jin1}.
We first couple the quarks to an
 external isovector-scalar field $S_V(x)$ which is described by adding
a term $\Delta {\cal L}$ to the usual QCD Lagrangian
\begin{equation}
\Delta{\cal L} \equiv - S_V(x) [\overline u (x) u(x)
-\overline d(x) d(x)]\; .
\label{lag}
\end{equation}
We can take $S_V(x)$ to be a constant $S_V$ since we are only
interested in the zero momentum transfer matrix element.  The
correlation functions of the nucleon current in QCD vacuum in the
presence of $S_V$ will be computed. The term linear in $S_V$ gives the
matrix element of interest.

\subsection{QCD sum rules for $\langle p|\overline{u}u-\overline{d}
d|p\rangle$}

Consider the correlation function of the proton interpolating field in
the presence of a {\it constant} external isovector-scalar field
$S_{\mbox{\tiny{\rm V}}}$ which can be taken to be arbitrarily
small
\begin{equation}
\Pi(S_{\mbox{\tiny{\rm V}}},q)\equiv i\int d^4{x}e^{iq\cdot x}
\langle 0|{{\rm T}[\eta_p(x)\overline{\eta}_p(0)]}\rangle_{S_
{\mbox{\tiny{\rm V}}}}\ ,
\label{corr}
\end{equation}
where $\eta_p$ is the proton interpolating field introduced in
Ref.~\cite{ioffe1}
\begin{equation}
\eta_p(x)=\epsilon_{abc}
\left[{u^T_a}(x)C\gamma_\mu u_b(x)\right]
\gamma_5\gamma^\mu d_c(x)\ ,
\label{eta}
\end{equation}
where $u_a(x)$ and $d_c(x)$ stand for the up and down quark fields,
$a,b$ and $c$ are the color indices, and $C=-C^T$ is the charge
conjugation matrix. Since $S_V$ is a scalar,
Lorentz covariance and parity allow one to decompose
$\Pi(S_{\mbox{\tiny{\rm V}}},q)$ into two distinct structures
\begin{equation}
\Pi(S_{\mbox{\tiny{\rm V}}},q)\equiv \Pi^1(S_{\mbox{\tiny{\rm V}}},q^2)+
\Pi^q(S_{\mbox{\tiny{\rm V}}},q^2)\rlap{/}{q}\ .
\end{equation}
To the first order in the external field $S_{\mbox{\tiny{\rm V}}}$,
the
two invariant functions can be written as
\begin{equation}
\Pi^i(S_{\mbox{\tiny{\rm V}}},q^2)=
\Pi^i_0(q^2)+S_{\mbox{\tiny{\rm V}}}\Pi^i_1(q^2)
\end{equation}
for $\{i=1,q\}$, where $\Pi^i_0$ are the invariant functions in the
absence of the external field which give rise to the mass sum rules
discussed extensively in
Refs.~\cite{ioffe1,belyaev1,leinweber1,reinders1}.
We are concerned here with the linear response to the external field
given by $\Pi^i_1(q^2)$.

To derive a QCD sum rule, one first carries out an OPE, which will
express $\Pi^i_1(q^2)$ in terms of various
vacuum correlators, and then matches it to an expansion in terms of
physical intermediate states. Now the external field contributes to
$\Pi({S_{\mbox{\tiny{\rm V}}}},q)$ in two ways: it couples directly to
the quark fields in the propagating nucleon current and it also
polarizes the QCD vacuum. The chiral condensates of up and down quarks
change as follows

\begin{eqnarray}
\langle\overline{u}u\rangle_{S_{\mbox{\tiny{\rm
 V}}}}&=&\langle\overline{u}u\rangle_{\mbox{\tiny{\rm 0}}}-\chi
 S_{\mbox{\tiny{\rm V}}}\langle\overline{u}u\rangle_{\mbox
{\tiny{\rm 0}}} ,
\label{uc}
\\*[7.2pt]
\langle\overline{d}d\rangle_{S_{\mbox{\tiny{\rm
 V}}}}&=&\langle\overline{d}d\rangle_{\mbox{\tiny{\rm 0}}}+\chi
S_{\mbox{\tiny{\rm V}}}\langle\overline{d}d\rangle_{\mbox{\tiny
{\rm 0}}} ,
\label{dc}
\end{eqnarray}
where $\langle \hat{O}\rangle_{\mbox{\tiny{\rm 0}}}\equiv \langle
0|\hat{O}|0\rangle$.
Using Eq.~(\ref{lag}) it is easy to see that $\chi$ is related to the
correlation function
\begin{equation}
\chi\langle\overline{u}u\rangle_{\mbox{\tiny{\rm 0}}}\equiv
{i\over 2}\int d^4x \langle{\rm
T}\{\overline{u}(x)u(x)-\overline{d}(x)d(x),
 \overline{u}(0)u(0)-\overline{d}(0)d(0)\}\rangle_{\mbox
{\tiny{\rm 0}}}\ .
\label{chi}
\end{equation}
Similarly the mixed  quark-gluon condensates change as follows
\begin{eqnarray}
\langle g_s\overline{u}\sigma\cdot {\cal G} u
\rangle_{S_{\mbox{\tiny{\rm V}}}}
&=&\langle g_s\overline{u}\sigma\cdot {\cal G} u\rangle_{\mbox{\tiny
{\rm 0}}}-\chi_m
S_{\mbox{\tiny{\rm V}}}\langle g_s\overline{u}\sigma\cdot {\cal G}
u\rangle_{\mbox
{\tiny{\rm 0}}} ,
\label{uqc}
\\*[7.2pt]
\langle g_s\overline{d}\sigma\cdot {\cal G} d
\rangle_{S_{\mbox{\tiny{\rm V}}}}
&=&\langle g_s\overline{d}\sigma\cdot {\cal G} d\rangle_{\mbox{\tiny
{\rm 0}}}+\chi_m
S_{\mbox{\tiny{\rm V}}}\langle g_s\overline{d}\sigma\cdot {\cal G}
d\rangle_{\mbox
{\tiny{\rm 0}}} ,
\label{dqc}
\end{eqnarray}
where $\sigma\cdot {\cal G}\equiv
\sigma_{\mu\nu}{\cal G}^{\mu\nu}$ with ${\cal G}^{\mu\nu}$ the gluon
field tensor and $\chi_{\mbox{\tiny m}}$, the susceptibility
corresponding to the quark-gluon mixed condensate, can also be
expressed in terms of a spectral representation.

Since isospin is a good symmetry for hadron matrix elements we have
assumed that the response of the up and down quarks is the same, apart
from sign, and that we can disregard the vchange in the gluon condensate
$\langle (\alpha_s/\pi)G^2\rangle$ due to the external isovector
field.

To calculate the Wilson coefficients of the OPE, we need the
coordinate-space quark propagators in the presence of the
external field and the vacuum condensates. To first order
in the external field $S_{\mbox{\tiny{\rm V}}}$, the propagators
in the fixed-point gauge \cite{fock1,schwinger1,jin4} take the
form
\begin{eqnarray}
\langle {\rm T}[u_{i}^a(x)\overline{u}_{j}^b(0)]\rangle_{S_
{\mbox{\tiny{\rm V}}}}&=
&{i\over 2\pi^2}\delta^{ab}{1\over
 x^4}\left[\hat{x}\right]_{ij}-\delta^{ab}{S_{\mbox{\tiny
{\rm V}}}\over 4\pi^2}{\delta_{ij}\over x^2}-
{1\over 12}\delta^{ab}\langle\overline{u}u\rangle_
{\mbox{\tiny{\rm 0}}}\delta_{ij}
\nonumber\\*[7.2pt]
& &
+{1\over 12}\delta^{ab}\chi
S_{\mbox{\tiny{\rm V}}}\langle\overline{u}u\rangle_{\mbox
{\tiny{\rm 0}}}\delta_{ij}
+\delta^{ab}{i S_{\mbox{\tiny{\rm V}}}\over
 48}\langle\overline{u}u\rangle_{\mbox{\tiny{\rm 0}}}
\left[\rlap{/}{x}\right]
\nonumber\\*[7.2pt]
& &
-\delta^{ab}{x^2\over 192 }\langle g_s\overline{u}\sigma\cdot
{\cal G} u
\rangle_{\mbox{\tiny{\rm 0}}}\delta_{ij}-\delta^{ab}\chi_
{\mbox{\tiny m}}S_{\mbox{\tiny{\rm V}}}{x^2\over 192 }\langle
 g_s\overline{u}\sigma\cdot {\cal G} u
\rangle_{\mbox{\tiny{\rm 0}}}\delta_{ij}
\nonumber\\*[7.2pt]
& &
+\delta^{ab}{i S_{\mbox{\tiny{\rm V}}}x^2\over 9\cdot 128}\langle
 g_s\overline{u}\sigma\cdot {\cal G}u
\rangle_{\mbox{\tiny{\rm 0}}}\left[\rlap{/}x\right]_{ij}
\nonumber\\*[7.2pt]
& &
-{ig_s\over 32\pi^2}\left(G_{\mu\nu}(0)\right)^{ab}
{1\over x^2}[\rlap{/}x\sigma^{\mu\nu}+\sigma^{\mu\nu}
\rlap{/}x]_{ij}+
\cdot\cdot\cdot\ ,
\label{propu}
\end{eqnarray}
\begin{eqnarray}
\langle {\rm T}[d_{i}^a(x)\overline{d}_{j}^b(0)]\rangle_
{S_{\mbox{\tiny{\rm V}}}}&=
&{i\over 2\pi^2}\delta^{ab}{1\over
 x^4}\left[\hat{x}\right]_{ij}+\delta^{ab}{S_{\mbox{
\tiny{\rm V}}}\over 4\pi^2}{\delta_{ij}\over x^2}-
{1\over 12}\delta^{ab}\langle\overline{d}d\rangle_
{\mbox{\tiny{\rm 0}}}\delta_{ij}
\nonumber\\*[7.2pt]
& &
-{1\over 12}\delta^{ab}\chi
S_{\mbox{\tiny{\rm V}}}\langle\overline{d}d\rangle_{\mbox{\tiny{\rm
 0}}}\delta_{ij}
-\delta^{ab}{i S_{\mbox{\tiny{\rm V}}}\over
 48}\langle\overline{d}d\rangle_{\mbox{\tiny{\rm 0}}}
\left[\rlap{/}{x}\right]
\nonumber\\*[7.2pt]
& &
-\delta^{ab}{x^2\over 192 }\langle g_s\overline{d}\sigma\cdot
{\cal G} d
\rangle_{\mbox{\tiny{\rm 0}}}\delta_{ij}+\delta^{ab}\chi_
{\mbox{\tiny m}}S_{\mbox{\tiny{\rm V}}}{x^2\over 192 }
\langle g_s\overline{d}\sigma\cdot {\cal G} d
\rangle_{\mbox{\tiny{\rm 0}}}\delta_{ij}
\nonumber\\*[7.2pt]
& &
-\delta^{ab}{i S_{\mbox{\tiny{\rm V}}}x^2\over 9\cdot 128}\langle
 g_s\overline{d}\sigma\cdot {\cal G}d
\rangle_{\mbox{\tiny{\rm 0}}}\left[\rlap{/}x\right]_{ij}
\nonumber\\*[7.2pt]
& &
-{ig_s\over 32\pi^2}\left(G_{\mu\nu}(0)\right)^{ab}
{1\over x^2}[\rlap{/}x\sigma^{\mu\nu}+\sigma^{\mu\nu}
\rlap{/}x]_{ij}+
\cdot\cdot\cdot\ .
\label{propd}
\end{eqnarray}
Here we are interested in terms linear in $S_{\mbox{\tiny{\rm V}}}$.
We can then disregard the current quark masses because
they make negligible contributions.

The correlation function
$\Pi(S_{\mbox{\tiny{\rm V}}},q)$ can be computed
using the Eqs.~(\ref{propu}) and (\ref{propd}) above.
We have computed the contributions
corresponding to the diagrams listed in Fig.~\ref{diag}.
The results of our calculations for the invariant functions
$\Pi^q_1$ and $\Pi^1_1$ are
\begin{eqnarray}
\Pi^q_1(q^2)&=&{1\over
 4\pi^2}\langle\overline{d}d\rangle_{\mbox{\tiny{\rm 0}}}
\ln(-q^2)+{4\over
3}\chi\langle\overline{u}u\rangle_{\mbox{\tiny{\rm 0}}}^2
{1\over q^2}-{1\over
 24\pi^2}\left(2\langle g_s\overline{u}\sigma\cdot {\cal G} u
\rangle_{\mbox{\tiny{\rm 0}}}-\langle g_s\overline{d}\sigma\cdot
{\cal G} d\rangle_{\mbox{\tiny{\rm 0}}}\right){1\over q^2}\nonumber
\\*[7.2pt]
& & + {\chi\over
6}\langle\overline{u}u\rangle_{\mbox{\tiny{\rm 0}}}
\langle g_s\overline{u}\sigma\cdot {\cal G} u
\rangle_{\mbox{\tiny{\rm 0}}}{1\over q^4} + {\chi_{\mbox{\tiny m}}\over
6}\langle\overline{u}u\rangle_{\mbox{\tiny{\rm 0}}}
\langle g_s\overline{u}\sigma\cdot {\cal G} u
\rangle_{\mbox{\tiny{\rm 0}}}{1\over q^4}\; ,
\label{q}
\\*[7.2pt]
\Pi^1_1(q^2)&=&{1\over 32
 \pi^4}(q^2)^2\ln(-q^2)+{\chi\over
4\pi^2}\langle\overline{d}d\rangle_{\mbox{\tiny{\rm 0}}}
q^2\ln(-q^2)-{2\over 3}\left(3
\langle\overline{u}u\rangle_{\mbox{\tiny{\rm
 0}}}\langle\overline{d}d\rangle_{\mbox{\tiny{\rm
 0}}}-2\langle\overline{u}u\rangle_{\mbox{\tiny{\rm 0}}}^2
\right){1\over q^2}\ .
\label{1}
\end{eqnarray}

We now turn to the phenomenological side of the sum rules, which
is obtained by expanding
$\Pi(S_{\mbox{\tiny{\rm V}}},q)$ in terms of
physical hadronic intermediate states. There are three types of
contributions. Firstly the matrix element of interest is contained
in the term
\begin{equation}
\langle 0|\eta_p|p\rangle\langle p|\overline{u}u-\overline{d}d|p
\rangle\langle p|\overline{\eta}_p
|0\rangle\ ,
\label{double}
\end{equation}
where the current $\overline{\eta}_p$ creates a proton that interacts
with the external field $S_{\mbox{\tiny{\rm V}}}$ and is then
annihilated by the current $\eta_p$. Defining
\begin{equation}
\langle 0|\eta_p|p\rangle=\lambda_p v\ ,
\end{equation}
where $\lambda_p$ denotes the coupling between
$\overline{\eta}_p(0)|0\rangle$ and the physical proton state and $v$
is the usual Dirac spinor ($\overline{v}v=2M_p$),
and introducing the notation
\begin{equation}
 H\equiv {\langle p|\overline{u}u - \overline{d}d |p \rangle\over 2 M_p}\; ,
\label{defH}
\end{equation}
one can write the
contribution of Eq.~(\ref{double}) to $\Pi^i_1(q^2)$
as
\begin{equation}
-\lambda_p^2{\rlap{/}q+M_p\over q^2-M_p^2}\hspace*{0.15cm} H
\hspace*{0.15cm}
{\rlap{/}q+M_p\over q^2-M_p^2}\ .
\label{dia_con}
\end{equation}
It is seen that the above term has a double pole at the nucleon mass.

The external field $S_{\mbox{\tiny{\rm V}}}$ can also
cause transition between
the proton and an excited state  which can have either positive or
negative parity relative to the proton. When the relative parity
is positive, the contribution can be written as
\begin{equation}
-\lambda_p\lambda_{p^*}{\rlap{/}q+M_p\over q^2-M_p^2}\hspace*{0.15cm}
H^*
\hspace*{0.15cm}
{\rlap{/}q-M^*\over q^2-{M^*}^2}\ .
\end{equation}
where now $H^*$ is the transition matrix element between $|p\rangle$
and $|p^* \rangle$. This term has a simple pole at the proton mass as
well as at the mass $M^*$ of the excited state.
It is easy to see that after a Borel transformation this contribution
is not exponentially damped as compared to the double pole contribution
Eq.~(\ref{dia_con}). Therefore, one has to retain these simple pole terms
in the analysis of the sum rules through the introduction of a
phenomenological parameter to be determined along with the diagonal
matrix element $H$ (see Refs.~\cite
{ioffe2,balitsky1,chiu1,belyaev2,chiu2,kolesinchenko1,belyaev3,jin1}).
The third type of
contributions comes from transitions involving only the excited states.
These are of course exponentially damped after a Borel transformation
and can be approximated in the usual manner
\cite{ioffe1,belyaev1,leinweber1,reinders1},
by equating them to the perturbative contributions starting from an
effective threshold.

Equating the OPE results Eqs.~(\ref{1}) and (\ref{q}) and  the
physical
intermediate state expansion discussed above and applying the Borel
transformation \cite{svz},
we obtain the following sum rules
\begin{eqnarray}
{M^2\over 4\pi^2}\langle\overline{q}q\rangle_{\mbox{\tiny{\rm 0}}}
E_0 L^{-4/9}+{4\over
3}\chi\langle\overline{q}q\rangle_{\mbox{\tiny{\rm 0}}}^2 L^{4/9}-
{1\over 24\pi^2}\langle g_s\overline{q}\sigma&\cdot&G q
\rangle_{\mbox{\tiny{\rm 0}}} L^{-8/9}  -
{\chi\over 6M^2}\langle\overline{q}q\rangle_{\mbox{\tiny{\rm 0}}}
\langle g_s\overline{q}\sigma\cdot G q
\rangle_{\mbox{\tiny{\rm 0}}}L^{-2/27}
\nonumber\\*[7.2pt]
-{\chi_m\over 6M^2}\langle\overline{q}q\rangle_{\mbox{\tiny{\rm 0}}}
\langle g_s\overline{q}\sigma\cdot G q
\rangle_{\mbox{\tiny{\rm 0}}}L^{-2/27}
&=&\left[2\lambda_p^2{M_p\over M^2}H
+A_q\right]e^{-M_p^2/M^2} \; ,
\label{sum_q}
\end{eqnarray}
\begin{eqnarray}
{M^6\over 16\pi^4}E_2L^{-8/9}+{\chi\over
4\pi^2}\langle\overline{q}q\rangle_{\mbox{\tiny{\rm 0}}}M^4E_1&-&{2
\over 3}\langle\overline{q}q\rangle_{\mbox{\tiny{\rm 0}}}^2
\nonumber\\*[7.2pt]
&=&\left[2\lambda_p^2{M_p^2\over M^2}H
+A_1\right]e^{-M_p^2/M^2} \; ,
\label{sum_1}
\end{eqnarray}
where $A_1$ and $A_q$ are the phenomenological parameters that
represent the sum over the contributions from all off-diagonal
transitions between the proton and the excited states. Here we have defined
$E_0\equiv 1-e^{-s_0/M^2}$, $E_1\equiv 1-e^{-s_0/M^2}\left({s_0\over
M^2}+1\right)$ and $E_2\equiv 1-e^{-s_0/M^2}\left({s_0\over 2M^4}
+{s_0\over M^2}+1\right)$, which account for the sum of the
contributions involving excited states only, where $s_0$ is an
effective continuum threshold.  In Eqs.~(\ref{sum_q}) and
(\ref{sum_1}), we have ignored the isospin breaking in the vacuum
condensates.
We have also taken into account the anomalous dimension
of the various operators through the factor
$L\equiv\ln(M^2/\Lambda_{\rm QCD}^2)/\ln(\mu^2/\Lambda_{\rm
QCD}^2)$\cite{svz,ioffe1}. We
take the renormalization scale $\mu$ and the QCD scale parameter
$\Lambda_{\rm QCD}$ to be $500\,\text{MeV}$ and $150\,\text{MeV}$.


\subsection{Estimate of $\chi$}

It is clear from Eq.~(\ref{chi}) that $\chi$ is determined
once the isovector-scalar two point function is known.
The latter has been studied by
Gasser and Leutwyler using chiral
perturbation theory to one loop \cite{gasser2}. They found
\begin{equation}
i\int d^4x \langle 0|{\rm T}\{\overline{u}(x)u(x)-\overline{d}
(x)d(x), \overline{u}(0)u(0)-\overline{d}(0)d(0)\}|0\rangle
= 8\left({m_\pi^2\over
m_u+m_d}\right)^2 h_3 \ .
\label{chi2}
\end{equation}
 Since two pions cannot form an isovector-scalar, the $|\pi \pi\rangle$
intermediate state does not contribute in Eq.~(\ref{chi2}).  The other
possible pseudoscalar two particle states are $| K \overline K\rangle$
and $|\eta \pi\rangle$ which means that an extension to $SU (3)$
flavor symmetry is necessary. This has been done by Gasser and
Leutwyler in Ref.~\cite{gasser3}. Using Eq.~(11.6) of
Ref.~\cite{gasser3}, and $\langle\overline{s}s\rangle_{\mbox{\tiny {\rm
0}}}/\langle\overline{q}q\rangle_{\mbox {\tiny{\rm 0}}}=0.8$, we get
$h_3\simeq -0.003$. Combining Eqs.~(\ref{chi}) and (\ref{chi2}), we
obtain
\begin{equation}
\chi = -4{m_\pi^2\over (m_u+m_d)f_\pi^2} h_3\simeq 2.2 \,
\text{GeV}^{-1} \; ,
\label{chiva}
\end{equation}
where we have used $(m_u+m_d)
\langle\overline{q}q\rangle_{\mbox{\tiny{\rm 0}}}=-m_\pi^2 f_\pi^2$,
and the experimental values $m_\pi=138\,\text{MeV}$,
$f_\pi=93\,\text{MeV}$, and a median value
$\langle\overline{q}q\rangle_{\mbox{\tiny{\rm 0}}}= -(240\text{MeV})^3$
which corresponds to $m_u+m_d=11.8\,\text{MeV}$.  Alternatively, $\chi$
can be determined as follows. The terms proportional to the current
quark masses in the QCD Lagrangian can be written as
\begin{equation}
{\cal L}_{\mbox{mass}} =-\hat{m}(\overline{u}u + \overline{d}d)+{1\over
 2}\delta m (\overline{u}u -
\overline{d}d)-m_s\overline{s}s-\cdots\ ,
\end{equation}
where $\hat{m}\equiv {1\over 2}(m_u+m_d)$ and the ellipses denote the
terms due to heavier quarks. Treating $\delta
 m(\overline{u}u-\overline{d}d)$ as a
source term one obtains using Eq.~(\ref{chi})
\begin{equation}
\chi \langle\overline{u}u\rangle_{\mbox{\tiny{\rm 0}}}={d \over
d \delta m}\langle\overline{u}u-\overline{d}d\rangle_{\mbox
{\tiny{\rm 0}}}\ .
\end{equation}
On the other hand, one can expand
$\langle\overline{u}u-\overline{d}d\rangle_{\mbox{\tiny {\rm 0}}}$ and
${d \over d \delta m} \langle\overline{u}u-\overline{d}d
\rangle_{\mbox{ \tiny{\rm 0}}}$ into the Taylor Series in $\delta m$.
Using $\langle\overline{u}u-\overline{d}d\rangle_{\mbox{\tiny
{\rm 0}}}|_{\delta m=0}=0$, we find
\begin{equation}
\langle\overline{u}u-\overline{d}d\rangle_{\mbox{\tiny{\rm 0}}}=
\chi\delta m
\langle\overline{u}u\rangle_{\mbox{\tiny{\rm 0}}}+ O[(\delta m)^2]\ ,
\end{equation}
which implies
\begin{equation}
\chi\delta m=-\gamma+O[(\delta m)^2]\ .
\label{chiper}
\end{equation}
Therefore, to the lowest order in $\delta m$, the susceptibility
 $\chi$ is determined by the ratio of the isospin breaking parameters
$\gamma$ and $\delta m$.
 The value of $\gamma$ has been estimated previously in various
approaches
\cite{gasser3,paver1,pascual1,bagan1,dominguez1,dominguez2,narison1},
with results ranging from $-1\times 10^{-2}$ to $-3\times 10^{-3}$.
Gasser and Leutwyler \cite{gasser2} have determined the ratio
	$\delta m /(m_u + m_d ) = 0.28 \pm 0.03 $.  Since we have used
a median value of 11.8 MeV for the sum of the up and down quark masses
we adopt a median value for $\delta m=3.3\,\text{MeV}$. Here we consider
the $\gamma $ values to be in the range $-1\times 10^{-2}$ to $-3\times
10^{-3}$, which, upon using Eq.~(\ref{chiper}), corresponds to
\begin{equation}
0.9\,\text{GeV}^{-1}\le\chi\le 3\,\text{GeV}^{-1}\ .
\label{chi-gamma}
\end{equation}

The susceptibility $\chi_{m}$ can also be determined using
a spectral representation for Eqs.~(\ref{uqc}) and (\ref{dqc}).
However, given the uncertainty in the value of $\chi$, we assume, in this
work, $\chi_{m}\simeq\chi$.

\subsection{Sum-rule analysis}

Defining $a\equiv-4\pi^2\langle\overline{q}q
\rangle_{\mbox{\tiny{\rm 0}}}$, $\tilde{\lambda}_p^2\equiv
32\pi^4\lambda_p^2$,
$m_0^2\equiv\langle\overline{q}g_s\sigma\cdot G q\rangle_
{\mbox{\tiny{\rm 0}}}/ \langle\overline{q}q
\rangle_{\mbox{\tiny{\rm 0}}}$, $\tilde{A}_q=(2\pi)^4  A_q$,
and $\tilde{A}_1=(2\pi)^4  A_1$, we can rewrite the sum rules
Eqs.~(\ref{sum_1}) and (\ref{sum_q}) as
\begin{eqnarray}
& &e^{M_p^2/M^2}\left[-M^4aE_0 L^{-4/9}+{4M^2\over 3}\chi
a^2 L^{4/9}+{M^2\over 6}m_0^2 a L^{-8/9}-{\chi\over3}a^2
m_0^2 L^{-2/27}\right]
\nonumber\\*[7.2pt]
& &\hspace*{2in}=\tilde\lambda_p^2 M_p H
+\tilde A_q M^2 \; ,
\label{fiq}
\end{eqnarray}
\begin{eqnarray}
& &e^{M_p^2/M^2}\left[M^8E_2L^{-8/9}-M^6\chi a E_1-{2\over 3} M^2
a^2\right]
\nonumber\\*[7.2pt]
& &\hspace*{2in}=\tilde\lambda_p^2 M_p^2 H
+\tilde A_1 M^2 \; .
\label{fi1}
\end{eqnarray}
To extract $H = \langle p|\overline{u}u-\overline{d}d|p\rangle /2M_p$,
$\tilde{A}_q$ and $\tilde{A}_1$ from the above sum rules, we use the
experimental value for the proton mass $M_p$ and extract
$\tilde{\lambda}_p^2$  from the proton {\it mass} sum rules.  Using
$a=0.55\,\text{GeV}^3 (m_u+m_d=11.8\,\text{MeV})$ and
$m_0^2=0.8\,\text{GeV}^2$, one finds from the best fit of the proton
mass sum rules that $\tilde{\lambda}_p^2=2.1\,\text{GeV}^6$
corresponding to $s_0=2.3\,\text{GeV}^2$\cite{ioffe2}.We use only the
first or the chiral-odd sum rule of Ioffe which is more accurate than
the second or the chiral-even sum rule.

First consider the sum rule Eq.~(\ref{fiq}), which is obtained from
$\Pi^q_1(q^2)$. For definiteness we take for $\chi$ the value
$\chi=2.2\,\text{GeV}^{-1}$ given in Eq.~(\ref{chiva}).  In
Fig.~\ref{sides1}, we plot the individual terms in the LHS of
Eq.~(\ref{fiq}) as well as their sum, as functions of $M^2$ in the
interval $0.8\leq M^2\leq 1.4\,\text{GeV}^2$. It can be seen that the
LHS follows a linear behavior in $M^2$ and we can match it with the
RHS. To find the best values for the constant and the coefficient of
the linear term in the RHS
 we follow the numerical optimization procedure used in
Refs.\cite{leinweber1,furnstahl1}.
We sample the sum rules in the fiducial region of $M^2$, where the
contributions from the highest-dimensional condensates included in the
sum rule remain small and the continuum contribution is controllable.
Here we choose $0.8\leq M^2\leq 1.4\,\text{GeV}^2$ as the optimization
region, which is identified by Ioffe and Smilga\cite{ioffe2} as the
fiducial region for the nucleon mass sum rules. To quantify the fit of
the left- and right-hand sides, we use the logarithmic measure
\begin{equation}
\delta(M^2)=\ln\left[{{\mbox{maximum}}\{LHS,RHS\}\over{\mbox
{minimum}}\{
LHS,RHS\}}\right]\ ,
\end{equation}
which is averaged over  $100$ points evenly spaced within the fiducial
region of $M^2$, where $LHS$ and $RHS$ denote the left- and right-hand
sides of the sum rules respectively. The sum-rule predictions are
obtained by minimizing $\delta$.  Using this procedure we obtain
\begin{equation}
 H \simeq 0.54\; , \; \tilde A_q
\simeq 0.29\,\text{GeV}^{5}\ .
\end{equation}
The RHS of Eq.~(\ref{fiq}) with these optimized values is also plotted in
Fig.2. One can see that the LHS and RHS have a very good overlap.

We now turn to the other sum rule Eq.~(\ref{fi1}), which is obtained
from $\Pi^1_1$.  In Fig.~\ref{sides2}, the individual terms in the LHS
as well as their sum are shown as functions of $M^2$ for
$\chi=2.2\,\text{GeV}^{-1}$.  The LHS is fairly linear in $M^2$ in the
interval $0.8 \leq M^2 \leq 1.4 \text{GeV}^{-2}$. The RHS of
Eq.~(\ref{fi1}), with the optimized values
\begin{equation}
H  \simeq 0.01\; , \; \tilde A_1
\simeq -1.82\,\text{GeV}^{6}\ ,
\end{equation}
is also shown in Fig.~\ref{sides2}.

It is worth  noting that in the sum rule Eq.~(\ref{fiq}) the double
pole term is more important than the single pole term, which is also
qualitatively evident from the near constancy of the LHS as a function
of $M^2$. In contrast, the single pole term in the sum rule
Eq.~(\ref{fi1}) clearly dominates, leading us to suspect that it is not
reliable to determine the double pole term in which we are interested.
This is confirmed by the following.

Let us  consider relaxing our tacit assumption that the continuum
threshold used in our external field sum rules should be the same as
the one occurring in Ioffe's mass sum rules. If $s_0$ is varied from
$2.3 \text{GeV}^2$ to $2\text{GeV}^2$ or $2.6 \text{GeV}^2$, the result
for the matrix element $H$ extracted from the sum rule Eq.~(\ref{fi1})
changes from $0.01$ to $-0.07$ ($s_0 =2\text{GeV}^2$) or $+0.08$
($s_0=2.6\text{GeV}^2$) while the result from the sum rule
Eq.~(\ref{fiq}) changes only from $0.54$ to $0.52$ or $0.56$.


It is clear from the above analysis that the chiral-odd sum rule
Eq.~(\ref{fiq}) is extremely stable, while  the chiral-even sum rule
Eq.~(\ref{fi1}) is not. So, in this paper we shall disregard the
results based on the sum rule Eq.~(\ref{fi1}) and consider  only the
results from the stable sum rule Eq.~(\ref{fiq}).

The fact that one of the sum rules works well, while the other fails,
is not peculiar to the problem under study. This pattern is seen also
in a study of the isoscalar-scalar matrix element as well as in the sum
rules for the matrix elements of electromagnet current and axial vector
current
\cite{ioffe2,balitsky1,chiu1,belyaev2,chiu2,kolesinchenko1,belyaev3,jin1}.
As discussed extensively in Ref.~\cite{chiu2}, the different asymptotic
behavior of  various sum rules can be traced to the fact that even and
odd parity states contribute with different sign and kinematical
factors. If chiral symmetry is realized in the Wigner-Weyl mode at high
energies, i.e., by parity doubling, it is possible to have either
cancellation or reinforcement between excited state contributions.
Irrespective of the exact manner in which these cancellations take
place, it is clear that the sum rule in which the continuum
contributions are weak is more reliable. This is the case for the
chiral-odd sum rule Eq.~(\ref{fiq}), where the continuum factor $E_0$
appears, to be contrasted with  the chiral-even sum rule
Eq.~(\ref{fi1}), where the factor $E_2$ occurs.

Let us now consider the effect of varying $\chi$ , which as we have
seen earlier is not precisely known. We find that the quality of the
overlap of the two sides of the chiral-odd sum rule remains good (as
measured by $\delta$) as we change the value of $\chi$, and gets better
as $\chi$ increases and worse as $\chi$ decreases. We also find that
the continuum contribution gets larger as $\chi$ decreases and smaller
as $\chi$ increases.
 For $\chi$ values in the range $1.4\,\text{GeV}^{-1}\le\chi\le
3.0\,\text{GeV}^{-1}$, we find
\begin{equation}
\ H =0.32 -0.76\; .
\end{equation}
For $\chi<1.4\,\text{GeV}^{-1}$, we find that the continuum
contribution is larger than $50\%$.  This implies that the sum rule is
dominated by continuum and the predictions are not reliable for $\chi$
values smaller than $1.4\,\text{GeV}^{-1}$.


\section{The neutron-proton mass difference}
\label{npd}

In this section we consider the relation between the matrix element
evaluated in the last section and the neutron-proton mass difference.
First let us disregard electromagnetism  and work to the first order in
the quark mass difference $m_d-m_u$.

Consider the quark mass term
in the QCD Hamiltonian density ${\cal H}_{\mbox{\tiny QCD}}$, as given by
\begin{eqnarray}
{\cal H}_{\mbox{\rm
mass}}&=&m_u\overline{u}u+m_d\overline{d}d+m_s\overline{s}s+
\cdots
\nonumber\\*[7.2pt]
&=&\hat{m}(\overline{u}u+\overline{d}d)-{1\over 2}\delta m
 (\overline{u}u-\overline{d}d)+m_s\overline{s}s+\cdots\ .
\label{xbreaking}
\end{eqnarray}
The isospin symmetry is explicitly broken by the term proportional
to $\delta m$. Using covariant normalization for the hadron state
labeled by $h$ and momentum $k$
\begin{equation}
\langle k^\prime, h \mid k, h \rangle~~=~~ (2\pi)^3  {k^0} \delta^{(3)}
(\vec k^\prime - \vec k)\ ,
\end{equation}
and regarding the $\delta m$ term in Eq.~(\ref{xbreaking}) as a
small parameter,
we can make a perturbation expansion and write the shift in the mass
$M_h$ of the hadron as \cite{gasser1}
\begin{equation}
\delta (M_h)={- \delta m\over 2} {\langle h|
(\overline u u - \overline d d)|h\rangle\over 2 M_h}\; .
\label{delm}
\end{equation}
The matrix element occurring in Eq.~(\ref{delm}) is to be computed at
$\delta m = 0$, which means isospin can be taken to be exact. We can
then write for the difference between the neutron and proton mass to
first order in $\delta m$ as
\begin{equation} (M_n-M_p)_q =\delta
m {\langle p | \overline u u - \overline d d | p \rangle\over 2 M_p}\; .
\end{equation}
The subscript $q$ in the left-hand side denotes the fact
that we are considering only the non-electromagnetic part of the mass
difference.

The effect of turning on electromagnetism is described by an
effective Lagrangian
\begin{equation}
{\cal L}_{\rm e.m.} ~~=~~ - {1 \over 2} e^2 \int d^4y
D(x-y) T j ^\mu (x) j_\mu (y) \; ,
\end{equation}
where $D(x) = [i 4 \pi^2 (z^2-i \epsilon)]^{-1}$~ is the
photon propagator and $j^\mu(x)$~ is the electromagnetic
current. To remove the divergence arising from electromagnetism
one must add counter terms, and these, of course, depend on
renormalization prescription. However, this dependence is extremely
weak. For example the change in the up quark mass due to a change
in the renormalization scale by a factor of two is less than 0.01 MeV.
 It is therefore
meaningful to separate the contribution from the quark mass difference,
from that due to  electromagnetism. The latter is estimated to be
\cite{gasser1}
\begin{equation}
(M_n-M_p)_{\mbox\tiny{\rm elec}}=-0.76\pm 0.30 \; .
\end{equation}
The experimental mass difference is 1.29 MeV, which
then gives
\begin{equation}
(M_n-M_p)_q^{\mbox\tiny{\rm
exp}}=2.05\pm 0.30 \; .
\end{equation}

In the last section we saw that uncertainty in our knowledge of $\chi$
leads to a corresponding uncertainty in our determination of
$H$.
For the $\chi$ value obtained from chiral
perturbation theory [see Eq.~(\ref{chiva})], we get
$(M_n-M_p)_q\simeq
1.8\,\text{MeV}$.

 For $\chi$ values in the range
$2.15\,\text{GeV}^{-1}\le\chi\le 2.80\,\text{GeV}^{-1}$, we find
\begin{equation}
1.75\text{MeV}\le (M_n-M_p)_q \le 2.35\text{MeV} \,,
\end{equation}
which is consistent
with experimental data. Smaller and larger values of $\chi$
outside the range considered above lead to correspondingly
smaller and larger values for
the neutron-proton mass difference.

\section{Discussion}
\label{discussion}

One of our main objectives in this paper has been to extract the proton
matrix element $H=\langle
p|\overline{u}u-\overline{d}d|p\rangle/2M_p$.  We have seen that the
chiral-odd sum rule Eq.~(\ref{fiq}) is reliable for determining this
matrix element. The major limiting factor has been the uncertainty in
the value of the susceptibility $\chi$.  A more accurate evaluation of
the two point function Eq.~(\ref{chi}) should reduce the uncertainty in
the value of $\chi$ and hence help to pin down the value of the matrix
element $H$.

We also saw that the non-electromagnetic part of the neutron-proton
mass difference is essentially given by the matrix element $H$
multiplied by the light quark mass difference $\delta m$.  If we use a
median value $\delta m  = 3.3$ MeV and $H \simeq 0.54$ as obtained
using a value of $\chi = 2.2 \text{GeV}^{-1}$ we get $(M_n-M_p)_q\simeq
1.8\,\text{MeV}$, which indeed has the right sign and magnitude. This
suggest that our approach to extract $H$ and the neutron-proton
mass difference is reliable. However, since $\chi$ and
$\delta m $ are not precisely known we cannot make a critical
comparison with data at present.

We now turn to a comparison of our method with those of earlier
authors. The nucleon mass was originally extracted by Ioffe
\cite{ioffe1} with a {\it combined} use of both the chiral-odd and
chiral-even mass sum rules.  It is necessary to use these two sum rules
together since one must eliminate the coupling constant $\lambda_N^2$.
Belyaev and Ioffe \cite{belyaev1} extended this method to determine the
mass splitting between hyperon and nucleon by treating the strange
quark mass as a perturbation. In addition to the mass shift, $M_Y -
M_N$, one must also take into account the change in the coupling
constant $\lambda_Y ^2 - \lambda_N ^2$  and the change in the continuum
threshold $s_0$. The authors of Refs.~\cite{adami1,eletsky1} used the same
procedure to determine the mass splittings within an isospin multiplet
by treating $m_d$ and $m_u$ as perturbation parameters. In
Ref.~\cite{hatsuda1} the neutron-proton mass was extracted directly
from  the difference between the neutron and proton mass sum rules, but
the continuum contributions were disregarded. In Ref.~\cite{adami1},
Adami {\it et.al.} retained continuum corrections but regarded $\gamma $
as a parameter to be determined by a fit to all isospin splittings in
the baryon octet. In Ref.~\cite{yang1}, apart from the perturbation due
to the quark masses, an attempt was made to incorporate the
electromagnetic contribution also phenomenologically in the sum rules.

The sum rules derived by us in Sec.~\ref{sumrule} can also be derived
directly from the mass sum rules.  Writing $m_u =\hat m - \delta m /2$,
$m_d =\hat m + \delta m /2$ and using  $\chi = -\gamma /\delta m $ one
can differentiate Eqs.~(16) and (17) of Ref.~\cite{yang1} with respect
to $\delta m$.  One can then identify our sum rules Eqs.~(\ref{fiq})
and (\ref{fi1}) with Eqs.~(29) and (30) of Ref.~\cite{yang1}. (An
assumption about the mixed chiral condensate, equivalent to our
assumption, $\chi_m = \chi$, was made in Ref.~\cite{yang1}).  This
coincidence between the sum rules is not surprising, since the quark
mass term in the QCD Lagrangian can also be regarded as a constant
external scalar field. The double pole term in the RHS of our sum rules
clearly arises from the simple pole term in the mass sum rules after
the differentiation of the proton mass with respect to $\delta m$.

We note that the term proportional to the mixed quark-gluon condensate
[the fourth term in the LHS of our Eq.~(\ref{fiq})] has not been
included in Ref.~\cite{adami1}. We have seen in our analysis of the sum
rules (see Fig.2) that this term is numerically significant. There is
also a minor discrepancy in the coefficient of the four-quark
condensate term between Ref.~\cite{adami1} and this work (or
Ref.~\cite{yang1}).  This discrepancy comes from the fact that the
authors of Ref.~\cite{adami1} directly used the $\Sigma$ and $\Xi$ mass
sum rules from Ref.~\cite{belyaev1}, where {\it not} all the quark mass
terms were taken into account.

It is now easy to see, as discussed earlier in Sec.~\ref{sumrule},
 why Eq.~(\ref{fiq}) is a better sum rule as compared to the
chiral-even sum rule Eq.~(\ref{fi1}). In the chiral-odd sum rule, the
perturbation due to the finite quark mass does not affect the leading
asymptotic behavior and hence when a differentiation with respect to
$\delta m$ (alternately when the neutron and proton sum rule
difference is taken) the leading asymptotic behavior reduces to a
smaller power of the Borel Mass $M^2$, which in turn means weaker
dependence on the continuum contribution. On the other hand, in the
chiral-even sum rule the introduction of quark mass leads to the term
$m_d M^6$ in the proton and $m_u M^6$ in the neutron sum rule.
Consequently in the difference the leading asymptotic behavior is now
$M^6$. In other words, the continuum contributions are enhanced. This
feature is also clearly reflected in our analysis in
Sec.~\ref{sumrule}. For the chiral-odd sum rule the double pole term
residue was stable when $s_0$ was varied while in the chiral-even sum
rule the corresponding residue was unstable. Further, in the
chiral-even sum rule the single pole term corresponding to transition
between proton and the excited states dominated over the double pole
term.
We would also like to point out that inclusion of instanton
contributions improves the chiral-even mass sum rule
\cite{hil}.  In the light of our observations above regarding the
relation of our sum rules to the mass sum rules it suggests that such
instanton contributions can also be significant in our sum rule
Eq.~(\ref{fi1}).

It is also worth emphasizing that in our work
two stages are involved. We first calculate a hadronic
matrix element $H$. This involves the susceptibility
$\chi$, which being essentially the ratio $-\gamma$/$\delta m$
is relatively insensitive to errors in our knowledge of $\delta m$.
The quark mass part of the neutron-proton mass difference was obtained
as the product $H \delta m $.

As a final remark we shall comment on the Nolen-Schiffer
anomaly \cite{nsa}. We saw in the last section that the empirical
neutron-proton mass difference can be written as
\begin{equation}
(M_n-M_p)^{\mbox\tiny{\rm exp}}~~=-0.76 + \delta m H \; .
\label{mnp}
\end{equation}
Now, it is well known that the matrix element of the axial vector current
is quenched \cite{ris} inside the nuclear medium by about 30 percent.
It may be reasonable to assume that the isovector-scalar
matrix element $\langle p| \overline u u - \overline d d
|p\rangle/2M_p$
is also quenched in a similar fashion.

Assuming then for example a 30 percent reduction in the value of $H$,
it follows form Eq.~(\ref{mnp}) that
the effective mass difference in the nuclear medium is
$(M_n-M_p)^{\mbox\tiny{\rm exp}}_{\mbox\tiny{\rm{med}}} \simeq
0.49\text{ MeV}$. Understanding the Nolen-Schiffer anomaly  is then
reduced to explain the quenching of $\langle p | \overline u u -
\overline d d | p \rangle/2M_p$ in nuclear medium. This can be handled
either by traditional nuclear structure calculations or again by use of
QCD sum rules as in the quenching of nucleon axial coupling
\cite{pp,dl}.


\acknowledgements

We thank T.~D.~Cohen and H.~Forkel for useful conversations,
and S.~L.~Adler for a useful correspondence. One of us (J.P.)
would like to thank D. Sen and A.~D. Patel for discussions.
X.J.acknowledges support from the National Science Foundation under
Grant No.\ PHY-9058487 and from the Department of Energy
under
Grant No.\ DE--FG05--93ER--40762.
M.N. acknowledges the warm hospitality and congenial
atmosphere
provided by the University of Maryland Nuclear Theory
Group and
support from FAPESP  BRAZIL.



\begin{figure}
\caption{Diagrams for the calculation of the Wilson coefficients of
the
correlation function. The solid, wavy
and dashed lines represent the quark, gluon, and external
field,
respectively.}
\label{diag}
\end{figure}
\begin{figure}
\caption{Borel mass dependence of the left-hand side (solid curve)
and
right-hand side (long-dashed curve) of Eq.~(\protect{\ref{fiq}}), with
$\chi=2.2\,\text{GeV}^{-1}$ and the optimized values $\langle
p|\overline{u}u-\overline{d}d|p\rangle/2M_p =0.54$ and $\tilde
A_q=0.29\,\text{GeV}^{5}$. The curves 1,~2,~3 and 4 correspond to the
first, second, third and fourth terms in the LHS of
Eq.~(\protect{\ref{fiq}}).}
\label{sides1}
\end{figure}
\begin{figure}
\caption{Borel mass dependence of the left-hand side (solid curve)
and
right-hand side (long-dashed curve) of Eq.~(\protect{\ref{fi1}}), with
$\chi=2.2\,\text{GeV}^{-1}$ and the optimized values $\langle
p|\overline{u}u-\overline{d}d|p\rangle/2M_p =0.01$ and $\tilde
A_1=-1.82\,\text{GeV}^{6}$. The curves 1,~2 and 3  correspond to the
first, second and third terms in the LHS of
Eq.~(\protect{\ref{fi1}}).}
\label{sides2}
\end{figure}



\begin{references}
%
\bibitem{svz}M.~A. Shifman, A.~I. Vainshtein, and V.~I.
Zakharov,
Nucl.\ Phys.\ {\bf B147}, 385 (1979);
{\bf B147}, 448 (1979);
{\bf B147}, 519 (1979).
%
\bibitem{ioffe2}B.~L. Ioffe and A.~V. Smilga,
Nucl.\ Phys.\ {\bf B232}, 109 (1984).
%
\bibitem{balitsky1}V.~M.~Balitsky and A.~V.~Yung, Phys.\
Lett.\
{\bf B129}, 328 (1983).
%
\bibitem{chiu1} C.~B.~Chiu, J.~Pasupathy, and S.~L.~Wilson,
Phys.\ Rev.\ {\bf D33}, 1961 (1986).
%
\bibitem{belyaev2}V.~M.~Belyaev and Y.~I.~Kogan, Phys.\ Lett.\
{\bf B136}, 27 (1984).
%
\bibitem{chiu2} C.~B.~Chiu, J.~Pasupathy, and S.~L.~Wilson,
Phys.\ Rev.\ {\bf D32}, 1786 (1985).
%
\bibitem{kolesinchenko1}A.~V.~Kolesnichenko, Sov. J. Nucl.
Phys. {\bf 39}, 968 (1984).
%
\bibitem{belyaev3}V.~M.~Belyaev and B.~Y.~Blok, Phys.\
Lett.\ {\bf B167}, 99 (1986).
%
\bibitem{jin1}X.~Jin,  M. Nielsen, and J. Pasupathy,
Phys.\ Lett. B {\bf 314}, 163 (1993).
%
\bibitem{gasser1}J. Gasser and H. Leutwyler, Phys.\ Rep.\
{\bf 87}, 77 (1982).
%
\bibitem{hatsuda1}T. Hatsuda, H. H{\o}gaasen, and M.
Prakash,
Phys.\ Rev.\ {\bf C42}, 2212 (1990).
%
\bibitem{yang1}K.-C. Yang, W.-Y.P. Hwang, E.M. Henley, and
L.S. Kisslinger, Phys.\ Rev.\ {\bf D48}, 3001 (1993).
%
\bibitem{adami1}C. Adami, E.~G. Drukarev, and B.~L. Ioffe,
Phys.\ Rev.\ {\bf D48}, 2304 (1993).
%
\bibitem{eletsky1}V.L. Eletsky and B.~L. Ioffe,
Phys.\ Rev.\ {\bf D48}, 1441 (1993).
%
\bibitem{ioffe1}B.~L. Ioffe,
Nucl.\ Phys.\ {\bf B188}, 317 (1981);
{\bf B191}, 591(E) (1981).
%
\bibitem{belyaev1}V.~M. Belyaev and B.~L. Ioffe,
Zh.\ Eksp.\ Teor.\ Fiz.\ {\bf 83}, 876 (1982)
[Sov.\ Phys.\ JETP {\bf 56}, 493 (1982)];
{\bf 84}, 1236 (1983)
[{\bf 57}, 716 (1983)].
%
\bibitem{leinweber1}D.~B. Leinweber,
Ann.\ Phys.\ (N.Y.) {\bf 198}, 203 (1990).
%
\bibitem{reinders1}For a review, see
L.~J. Reinders, H. Rubinstein, and S. Yazaki,
Phys.\ Rep.\ {\bf 127}, 1 (1985), and references therein.
%
%
\bibitem{fock1}V. Fock,
Physikalische Zeitschrift der Sowjetunion {\bf 12}, 404
(1937).
%
\bibitem{schwinger1}J. Schwinger,
{\it Particles, Sources, and Fields}
(Addison-Wesley, Reading, 1970), Vol.~I.
%
\bibitem{jin4}X. Jin, T.~D. Cohen, R.~J. Furnstahl, and D.~K.
Griegel,
Phys.\ Rev.\ C {\bf 47}, 2882 (1993), and references therein.
%
\bibitem{gasser2}J.~Gasser and H.~Leutwyler, Ann.\ Phys.\
{\bf 158},
142 (1984).
%
\bibitem{gasser3}J.~Gasser and H.~Leutwyler, Nucl.\ Phys.\
{\bf B250},
465 (1985).
%
\bibitem{paver1}N. Paver, Riazzudin, and M. D. Scadron, Phys.\
Lett.\ {\bf B197}, 430(1987).
%
\bibitem{pascual1}P. Pascual and R. Tarrach, Phys.\ Lett.\
{\bf 116B}, 443(1982).
%
\bibitem{bagan1}E. Bagan {\bf et al.}, Phys.\ Lett.\ {\bf 135B},
463(1984).
%
\bibitem{dominguez1}C. A. Dominguez and M. Loewe, Phys.\ Rev.\
{\bf D 31}, 2930(1985).
%
\bibitem{dominguez2}C. A. Dominguez and E. de Rafael, Ann.\
Phys.\ (N.Y.) {\bf 174}, 372(1987).
%
\bibitem{narison1}S. Narison, Rev.\ Nuovo Cimento {\bf 10},
1(1987);
{\it QCD Spectral Sum Rules}, World Scientific Lecture Notes
in Physics Vol. 26(World Scientific, Singapore, 1989).
%
%
\bibitem{furnstahl1}R.~J. Furnstahl, D.~K. Griegel, and T.~D.
Cohen,
Phys.\ Rev.\ C {\bf 46}, 1507 (1992);
X.~Jin, M.~Nielsen, T.~D. Cohen, R.~J. Furnstahl,
and D.~K. Griegel, Phys.\ Rev.\ C {\bf 49}, 464 (1994);
X.~Jin and R.~J. Furnstahl, Phys.\ Rev.\ C
{\bf 49}, 1190 (1994).
%
\bibitem{hil}H. Forkel and M.~K. Banerjee, Phys.\ Rev.\ Lett. {\bf 71},
484(1993).
%
%
\bibitem{nsa}J.~A. Nolen and J.~P. Schiffer, Ann. Rev. Nuc. Sci. {\bf 19},
414 (1969).
%
\bibitem{ris}D.~O. Riska and K. Tsushima, Nucl. Phys. {\bf A553},
684c (1993).

%
\bibitem{pp}R. Parthasarathy and J. Pasupathy, Phys.\ Rev.\ C {\bf 37},
2140 (1988).
%
\bibitem{dl}E.~G. Drukarev and E.~M. Levin,
Prog.\ Part.\ Nucl.\ Phys.\ {\bf 27}, 77 (1991);
Nucl.\ Phys.\ {\bf A511}, 679 (1990);
{\bf A516}, 715(E) (1990).
%
\end{references}
\end{document}